\documentclass[a4paper,twoside]{article}
\newcommand{\affil}[1]{$^{\rm #1}$}
\usepackage{graphicx}

\date{} 
\title{\large\bf\flushleft The Effect of the Circumstellar Material on the Light Curves of Eclipsing Binary Systems}
\author{\parbox{\textwidth}{\flushleft
\vspace{-0.5cm}
{\it S.M.R. Ghoreyshi\affil{a,b}, J. Ghanbari\affil{a,c}, and F. Salehi\affil{c}}\\
\vspace{0.4cm}
{\small \affil{a}\,Department of Physics, School of Sciences,
Ferdowsi University of Mashhad, Mashhad,Iran}\\
{\small \,m.ghoreyshi@crya.unam.mx} \\
{\small \affil{b}\,Centro de Radioastronomia y Astrofisica,
Universidad Nacional Autonoma de Mexico,
Apdo. Postal 72-3 (Xangari), Morelia, Michoacan 58089, Mexico}\\
{\small \affil{c}\,Department of Physics, Khayyam Institute of Higher Education, Mashhad, Iran}\\
 {\small \,ghanbari@ferdowsi.um.ac.ir} \\
 {\small \,f.salehi@khayyam.ac.ir}}}
\begin{document}
\twocolumn[
\begin{changemargin}{.8cm}{.5cm}
\begin{minipage}{.9\textwidth}
\vspace{-1cm}
\maketitle
%
%
\small{\bf Abstract:} This study inspects the influence of various
effects and free parameters of the accretion disc and circumstellar
material on the emerging light curve of eclipsing binary systems
that have a circumstellar disc, by using the SHELLSPEC code. The
results indicate that some of the parameters namely the temperature
and inclination of the disc, spot, jet, stream and shell
significantly affect on the emerging light curve while some other
parameters namely the exponent of the power-law behavior of the
density of the disc, microturbulence, inner and outer radius of the
disc don't noticeably affect on the emerging light curve. An
application to the Algol-type eclipsing binary system AV Del and an
accretion disc model for the system using the SHELLSPEC code is
included.


\medskip
\medskip
\end{minipage}
\end{changemargin}
]
\small

\section{Introduction}
In recent years the accretion disc is studied extensively. There are
many astrophysical systems that the accretion process has an
important role in their structure and evolution. The accretion discs
are seen in AGNs, protoplanet clouds and binary systems. The
accretion process usually occurs in the binary system, when one of
companions fills its Roche lobe. Another companion can be a black
hole, a neutron star, a white dwarf or a typical massive star. One
of more current type of binary system in which mass transfer are
seen between companions is the Algol-type binary system. The
formation and properties of the accretion disc and circumstellar
material in the Algol-type systems are not very well understood.

Various complicated computer codes are presented for calculating and
reversing light curves or spectra of binary stars with various
shapes or geometry including the Roche model (Lucy 1968; Wilson \&
Devinney 1971; Mochnacki \& Doughty 1972; Rucinski 1973; Hill 1979;
Zhang et al. 1986; Djurasevic 1992; Drechsel et al. 1994; Vink\'{o}
et al. 1996; Hadrava 1997; Bradstreet \& Steelman 2002; Pribulla
2004). In the codes, the stars are assumed to be nontransparent and
stripped of any circumstellar matter. An interesting approach for
studying extended semitransparent atmospheres of some eclipsing
binary components was developed by Cherepashchuk et al. (1984).
Often the three-dimensional model of the circumstellar matter
(behavior of state quantities and velocity field) is known or
expected as a result of hydrodynamic simulations or observational
constraints (see, e.g., Karetnikov et al. 1995; Richards \& Ratliff
1998). Unfortunately, three-dimensional non-local thermodynamic
equilibrium (NLTE) calculations and spectral synthesis including
complex hydrodynamics are difficult to carry out, so one alternative
has been to perform a simple volume integration of emissivity, which
is often too oversimplified for the particular problem. On the other
hand, highly sophisticated model atmospheres and spectrum synthesis
codes have been developed assuming NLTE and plane-parallel
atmospheres of hot stars (Hubeny 1988; Hubeny \& Lanz 1992, 1995;
Hubeny et al. 1994), spherically symmetric atmospheres (Kub\'{a}t
2001), or stellar winds (Krtic¡ka \& Kub\'{a}t 2002). There are also
sophisticated stationary planeparallel line-blanketed model
atmospheres and spectrum synthesis codes for a large variety of
stars assuming LTE (Kurucz 1993a, 1993b; Smith \& Dworetsky 1988;
Piskunov 1992; and many others). However, these are very specialized
codes and their main purpose is to calculate the spectrum emerging
from a stellar atmosphere. It is difficult to apply them to the
various cases outlined above. An exception is the special case of
circumstellar matter in the form of accretion discs in cataclysmic
variables (CVs). In such a case, the disc is approximated by a set
of geometrically thin, but optically thick, static local atmospheres
and the output radiation is a sum of properly Doppler-shifted local
emerging intensities (Orosz \& Wade 2003; Wade \& Hubeny 1998; la
Dous 1989). However, the case of optically thin discs or accretion
disc winds, the Sobolev approximation is used (Proga et al. 2002;
Long \& Knigge 2002; Rybicki \& Hummer 1983). Linnell \& Hubeny
(1996) developed a package of codes that can calculate light curves
or spectra of interacting binary stars including an optically thick
(nontransparent) disc. However, it is often necessary to solve the
radiative transfer in a moving disc at least along the line of
sight, as demonstrated by Horne \& Marsh (1986). Nevertheless, their
calculations assumed a linear shear and neglected stimulated
emission, continuum opacity, and scattering, as well as a central
star.

Budaj et al. (2005) attempted to bridge the gap in the previously
mentioned approaches by using their code SHELLSPEC. It is a tool
that solves in LTE the simple radiative transfer along the line of
sight in an optional optically thin three-dimensional moving medium.
Optional transparent (or nontransparent) objects such as a spot,
disc, stream, jet, shell, or stars, as well as an empty space, may
be defined (or embedded) in three dimensions and their composite
synthetic spectrum calculated. The stars may have the Roche geometry
and known intrinsic spectra. The code is quite a multipurpose,
independent, and flexible tool that can also calculate a light curve
or a trailing spectrogram when necessary, or it can be used to study
various objects or effects. Also, they performed investigations in
order to study the underlying physics of the structures of the
Algol-type eclipsing binaries, by using the SHELLSPEC code. They
calculated synthetic H$\alpha$ spectra of TT Hya, which is an
Algol-type eclipsing binary with an accretion disc. They studied the
influence of various effects and free parameters of the disc on the
emerging spectra. Also, Ghoreyshi et al. (2011) used the SHELLSPEC
code to solve the light curve of Algol-type eclipsing binary system
AV Del. But a complete study about the effect of these parameters on
the light curve is not published.

AV Del ($\alpha_{2000}$ = $20^h45^m31^s.47$, $\delta_{2000}$ =
$11^\circ10'26''.4$, \emph{V}=11.8) is a rather neglected variable
star which is discovered by Hoffmeister (1935) and is given the
original designation 184.1930 to it. It is a suspected member of the
cool Algols (Popper, 1996) and was reported by the author as a
double-lined system with a period of 3.85 days and a mean spectral
type of G5, although other sources have classified it as F8 (e.g.,
Halbedel 1984). The rare class of cool Algol systems are
distinguished from the classical Algol systems in that the
mass-gaining component is a late-type star rather than a B- or A-
type star. Mader et al. (2005) presented a new spectroscopic and
\emph{BVRI} photometric observations of AV Del. They determined the
radial-velocities for both components, using the two-dimensional
cross-correlation technique TODCOR (Zucker \& Mazeh 1994). To
present a detailed radial velocity and light curve analysis of the
optical data, Mader et al. (2005) combined all light curve data with
the spectroscopic observations and analyzed them using the
Wilson-Devinney (WD) code (Wilson \& Devinney 1971; Wilson 1979,
1990). They showed the system to be most likely semi-detached, with
the less massive and cooler star filling its Roche lobe.

The paper tried to explore in more detail the effects of various
input parameters on the light curve of a typical eclipsing binary
system which is an Algol-type kind, by using the SHELLSPEC code
(section 2).

In section 3 we explain the simulation results. In section 4 at
first we discuss the effects of various input parameters on the
light curve of a typical Algol-type eclipsing binary system and then
show how the SHELLSPEC code describes the light curve of the really
Algol-type eclipsing binary system AV Del and summarize the given
results in section 5.

\section{Code Description}

When a star (of a binary star) in a circular orbit expands to fill
and then overfill its Roche lobe, it will start to lose mass to its
companion. Matter that leaves the surface of one component of a
binary star can be partly or wholly accreted by the companion. The
matter can directly strike the gainer, or it can be stored in an
accretion disc around it, and a part of the matter may leave the
system.

In common situation, the matter that feeds the disc flows from the
companion star through the inner Lagrangian point that connects the
two Roche lobes in the binary system. The matter thus leaves the
mass-losing star in a rather thin stream in the orbital plane of two
stars. In reality, the matter may spray in a messier fashion, but
most of the matter remains in the orbital plane. If the two stars
were stationary, this matter would flow from one star directly along
the line connecting the centers of the two stars and strike the
mass-gaining star. In a binary star system, however, the stars are
constantly moving in orbit, so the mass-gaining star is a moving
target. The matter may leave the mass-losing star headed for the
other star, but because the other star moves along in its orbit, the
transferred matter cannot fall directly onto the mass-gaining star.

The gravitational domain of the mass-gaining star captures the
matter, however, so the stream circles around and collides with the
incoming stream. As this process continues, the flow of
self-interacting matter will form first a ring and then a disc
(Wheeler, 2007).

Accretion discs often show P-Cyg type absorption lines that may be
caused by coronal heating above and below the disc. A tenuous wind
may be driven away from the disc, somewhat like a stellar wind
(Eggleton, 2006). Moreover, a thin shell of gas may surround the
primary star that usually because of the stellar wind of the primary
expands.

The interaction of the mass stream with the disc (in the place where
the gas stream from the secondary strikes the edge of the disc)
causes a bright hot spot which is seen as additional light in the
light curve.

Many objects that involve accretion discs are also seen to be
accompanied by bipolar jets apparently emerging from the central
region normal to the disc. Jets are also seen in $\emph{active
galactic nuclei}$ (AGN), where it is believed that a central massive
black hole is accreting neighboring material.

The appearance to an observer varies greatly as determined by the
observer's location relative to the jet direction. Stellar objects
also exhibit accretion discs and jets such as protostars and
x-ray-emitting black hole binary systems. Collimated polar jets,
taking angular momentum away from the disc, allow most disc matter
to fall onto the compact object.

Protostars are embryonic stars in their formative stages before
setting on the main sequence. The black hole binaries are sometimes
called microquasars because they exhibit relativistic jets similar
to those of extragalactic quasars. In the binary accreting systems,
the accreted matter is provided by a normal gaseous companion star.
The masses of the black hole in microquasars are thought to be of
order 10 M$_\odot$ (Bradt, 2008).

All such effects can play an important role in the evolution of the
binary systems. Therefore, studying and deliberating of the
existence and the structure of the effects can give us good insights
about the interacting binary systems.

The Algol-type binaries are interacting systems consisting of a hot,
usually more massive, main-sequence star and a cool giant or
subgiant that fills its Roche lobe. The secondary is losing the mass
to its companion, and we are observing the circumstellar material in
the form of a gas stream, a jet, a shell, an accretion annulus, or
an accretion disc. The most recent paper that summarizes our current
knowledge of the Algols is the work by Richards \& Albright (1999).

We have used the SHELLSPEC code that is to our knowledge the only
tool that can produce the synthetic light curves of interacting
binaries, taking into account optically thin moving circumstellar
matter. The code was applied to the typical Algol-type eclipsing
binary which has an accretion disc. Synthetic light curve of the
system were calculated by varying the parameters.

The SHELLSPEC code is designed to solve simple radiative transfer
along the line of sight in three-dimensional moving media. The
scattered light from a central object is taken into account assuming
an optically thin environment (Budaj \& Richards 2004). Output
intensities are then integrated through the two-dimensional
projection surface of a three-dimensional object. The assumptions of
the code include LTE and optical known state quantities and velocity
fields in three dimensions (Budaj \& Richards 2004). In the code,
both objects the primary and the companion (secondary), may be
shaped according to Roche model used for detached, semi-detached or
contact systems.

Light curve changes were studied considering these parameters
varying: geometrical shape of the disc, half thickness of the disc
cylinder, the exponent of the power-law behavior of the densities
$\rho\sim$ $r^\eta$ in the disc, effective temperature of the disc,
inclination of the disc, microturbulence, inner radius of the disc,
outer radius of the disc and as well as existing of spot, jet,
stream and shell.


\section{Light Curve Changes by Varying Model Parameters}

Now, the purpose is to study in more detail the influence of the
effects on the light curve of a typical eclipsing binary system
which is an Algol-type one. At first, a typical Algol-type eclipsing
binary was considered in which the primary star was surrounded by an
accretion disc. The primary was treated as a rotating solid sphere
while the surface of the secondary was modeled in terms of the Roche
model. The primary has the following characteristics: the mean
temperature $T_1=9800$ K, the mass $M_1=2.6$ M$_\odot$ and the
radius $R_1=1.9$ R$_\odot$. Also, the secondary has the following
characteristics: the mean temperature $T_2=4600$ K, the mass
$M_2=0.6$ M$_\odot$ and the radius $R_2=5.9$ R$_\odot$. The
considered disc is an optically thick Keplerian disc that has the
following characteristics: inner radius $R_{\rm in}=2.0$ R$_\odot$,
outer radius $R_{\rm out}=9.0$ R$_\odot$, effective temperature
$T_{\rm disc}=7000$ K and half of the thickness of the disc cylinder
$\alpha=1.0$ R$_\odot$. The electron number density of the disc was
set to $n_e=2\times10^{10}$ cm$^{-3}$, density
$\rho=3.3\times10^{-14}$ gcm$^{-3}$, microturbulence $\upsilon_{\rm
trb}=30$ kms$^{-1}$ and inclination $i=82^\circ.84$.

The parameters were allowed to vary one by one and the others were
kept fixed. Therefore,our calculations illustrate the effect of
varying just one free parameter at a time while keeping all the
others fixed. In many cases this may not represent the real behavior
of the system, since some quantities may be more or less
interlocked, such as temperatures and electron number densities.
Nevertheless, the study give us a generally realization about the
behavior of system when the free parameters vary.

Figure 1 displays the effect of geometrical shape of the disc. Here,
three profiles are compared: downer profile was obtained for the
slab-shaped disc with the parameters listed earlier. The middle
profile was obtained for the shape of a rotating wedge with the
wedge angle $\alpha=25^\circ$. It is clear that both profiles are
similar, except that the rotating wedge shape produces luminosity of
the system more strength than the slab disc model because the
rotating wedge shape disc has Keplerian emission surface than the
slab disc model. The upper profile was obtained for the case of
without disc. As the considered disc has the less effective
temperature than the primary, therefore the overall luminosity
increases when there is not the disc. Of course if the inclination
decreases, the emission surface of the disc increases and the
overall luminosity of the system increases. Also, it is clear that
the main difference between three profiles is out of the primary
eclipse.

Figure 2 clarifies the effect of increasing the vertical
half-thickness of the disc from 0.25 to 3.5 R$_\odot$ for the
slab-shape disc model. The luminosity is weakest at about 1.5
R$_\odot$. The luminosity of the system decreases from 0.25 to 1.5
R$_\odot$ since the accretion disc which has the less temperature
than the primary star, creates an eclipse (similar to the
secondary). Increasing the vertical half-thickness from 1.5 to 3.5
R$_\odot$ increases the luminosity of the system since increasing
the emitting space volume dominates the effect of the eclipse. It is
clear that the main difference between the profiles is out of the
primary eclipse.

Figure 3 indicates the effect of increasing the vertical
half-thickness of the disc from $10^\circ$ to $30^\circ$ for the
rotating wedge disc model. The luminosity of the system increases as
the emitting space volume increases. The wedge angle
$\alpha=25^\circ$ was considered, therefore the half-thickness of
the disc near the surface of the primary is about 0.9 R$_\odot$ and
the half-thickness of the disc at the outer limb of the disc is
about 4 R$_\odot$. Thus, similar to the slab disc model, in the
current shape increasing of the emitting space volume dominates the
effect of the eclipse by the disc.

Figure 4 (top left and top right) displays the effect of changing
the outer radius of the disc from 5 to 11 $R_\odot$. For $R_{\rm
out}=7$ R$_\odot$, the luminosity of the system has the minimum
amount. The luminosity of the system decreases from 5 to 7
R$_\odot$, and increases from 7 to 11 R$_\odot$. The evidence of
this showing is similar to the case of the vertical half-thickness
parameter. Increasing the outer radius from 5 to 7 R$_\odot$, the
disc plays an obstructive role between us and the light of the
primary star. Then, increasing toward higher radius, the influence
of increasing of the emitting space volume overcomes the influence
of eclipse, therefore the overall luminosity increases.
Nevertheless, the difference between the profiles for 5 to 11
R$_\odot$ is not very noticeable, especially for 5 to 7 R$_\odot$.

The bottom part of Fig. 4 displays the effect of increasing the
inner radius of the disc from 2 to 8 R$_\odot$. Increasing from 2 to
3 R$_\odot$, the luminosity decreases since the emitting space
volume decreases. Increasing from 3 to 8 R$_\odot$, the luminosity
increases since the less volume of the disc lies between the primary
star and us. For $R_{\rm in}=8$ R$_\odot$ we have an accretion
annulus instead of the accretion disc. Here again the difference
between the profiles is not very noticeable and is out of the
primary eclipse.

Figure 5 illustrates the light curve if the disc was viewed from
different inclination angles. Starting from edge-on, $i=90^\circ$,
the system has the least luminosity and the deepest primary eclipse.
As the inclination decreases and one begins to see the disc pole-on,
the luminosity increases and the shape of the light curve noticeably
changes since that part of the surface of the system which is toward
us, increases. At the inclination angle $i=60^\circ$ the depth of
the primary and secondary eclipse are comparable. In fact, the depth
of the primary eclipse extremely decreases and the depth of the
secondary eclipse slightly increases. At the inclination angle
$i=30^\circ$, the depth of the primary and secondary eclipses are
equal approximately and their expanse extremely increase.

Figure 6 displays the effect of varying the temperature of the disc
over the interval 5000-12000 K. The luminosity is the weakest at
about 6000 K and increases toward higher temperatures since the
temperature of the disc will be comparable with that of the primary.
It is clear that increasing of the temperature does not affect on
the shape of the light curve but increases or decreases the
luminosity out of the primary eclipse.

If the density and electron number density be inhomogeneous and have
a power-law dependence on the distance from the star, namely,
$\rho\sim r^\eta$, Figure 7 clarifies the effect of varying the
exponent $\eta$ from 0 to -2. The parameter noticeably affects on
the emerging spectrum (Budaj et al. 2005) but the light curve is not
so sensitive to the density. It is obvious from the figure that
difference between profiles for $\eta=0$ and $\eta=-2$ is not
noticeable but for $\eta=-1$ there is a more difference from
$\eta=0$ and $\eta=-2$ and the luminosity for $\eta=0$ and $\eta=-2$
is more than $\eta=-1$.

From comparing densities for $\eta=-1$ and $\eta=-2$, we find that
density at the outer region of the disc is less than its inner
region so that it is tended lesser if it is considered for eta=-2.
Therefore we expect more luminosity for binary system in latter case
rather than the former. Since less density is caused less effect on
emerging the light of the star via attracting or scattering. Of
course, we must notice that the inclination of disc can influence
the emission surface of the disc and so changes total luminosity of
system. Also, the density is fixed and independent of $r$ for
$\eta=0$. In this case, we receive more luminosity for binary system
than the former case. It can involve same corporation in attracting
or scattering function for all formed elements of the disc. In
results, change of total luminosity isn't very noticeable.

Figure 8 shows the effect of increasing the microturbulence of the
disc from 30 to 90 kms$^{-1}$. Increasing the microturbulence of the
disc, the overall luminosity of the system increases, too. Like as
most of earlier cases, the changes in the profile are seen in out of
the primary eclipse.

Figure 9 (top left) illustrates the influence of existence of the
spot on the light curve. It is clear that the spot can noticeably
affect on the luminosity of the system. Main difference is in the
primary eclipse. In the figure a spot was considered which has the
effective temperature $T_{\rm spot}=4400$ K and radius $R_{\rm
spot}=1$ R$_\odot$. Therefore, the spot causes increasing of the
luminosity of the system. Amount of difference between profiles
depends on the free parameters of the spot.

Figure 9 (top right) displays the influence of existence of the jet
on the light curve. It was considered that the jet has the following
characteristics: angle half-width of the jet cones $\theta=
10^\circ$, inner radius boundaries of the jet cones $r_{\rm in}=1.0$
R$_\odot$, outer radius boundaries of the jet cones $r_{\rm
out}=5.0$ R$_\odot$ and radial (expanding) velocity of the jet
$V_{\rm j}=360$ kms$^{-1}$. As it is shown in Figure 5, the jet
noticeably changes the shape of the light curve and increases the
overall luminosity of the system. Existence of the jet causes
decreasing the depth of the primary eclipse and increasing the depth
of the secondary eclipse. Also, the shape of the light curve changes
the interval phases 0.1 to 0.4 and 0.6 to 0.9. In the intervals two
wings are appeared. At the primary eclipse the jet cannot be seen
totally because the secondary star covers a part of the jet but at
the interval phases 0.1 to 0.4 and 0.6 to 0.9 the jet can be seen
quietly. Also, the secondary star is not in eclipse, therefore two
wings are appeared. Possibly, it is not suitable that the effect of
the existence of the jet in the Algol-type binary systems was
considered since usually one cannot see the jet in these systems.
Perhaps, it is better that the influence of the jet in the binary
systems which include white dwarf, neutron star or black hole would
be investigated, but the aim of the study was to display the effect
of the jet on the shape of the light curve of Algol-type systems,
even if it is not so realistic.

Figure 9 (bottom left) illustrates the effect of existence of the
stream on the light curve. The considered stream has the following
characteristics: stream velocity at the beginning of the stream
$V_1=250$ kms$^{-1}$, stream velocity at the end of the stream
$V_2=530$ kms$^{-1}$ and radius of the stream $r_{\rm s}=1$
R$_\odot$. It is clear that the stream did not change the depth or
the expense of the primary and secondary eclipse and did not change
the luminosity of the system but it causes the appearance of two
wings at the first and second quadrature of the light curve.

Figure 9 (bottom right) displays the effect of existence of the
shell around the system. The considered shell has the following
characteristics: velocity of the uniformly expanding shell $V_{\rm
sh}=40$ kms$^{-1}$ and mean temperature of the shell $T_{\rm
sh}=9500$ K. It is clear that the shell extremely increases the
luminosity of the system. Also, the shell causes decreasing the
depth of the primary eclipse.

The purpose of the paper was not to study the effect of the free
parameters of the spot, jet, stream and shell on the light curve of
the system. Nevertheless, the effect of existence of the these
objects on the light curve of the system was studied.

Now that we understand the influence of various parameters on the
emerging light curve, we can proceed to fit the real light curve of
AV Del. We used the photometric data reported by Mader et al (2005)
for analyzing via the SHELLSPEC code. A detailed study of these data
published in Ghoreyshi et al. (2011) in which analysis the
photometric and radial-velocity data of the system by using both
Wilson-Devinney code (Wilson \& Devinney 1971; Wilson 1979, 1990)
and the SHELLSPEC code performed.

We performed some sophisticated calculations of the light curve in
which both stars and a disc were considered. The primary was treated
as a rotating solid sphere, while the surface of the secondary was
modeled in terms of the Roche model. First, the code was run
assuming there was no third body but it failed to obtain a good fit
to the observed data. Then a disc only was included as a third body.
On the basis of the configuration, agreement between the observed
and the theoretical light curve was very good, although the paper
tried to examine the effect of other bodies (spot, jet, stream and
shell) on the light curve. Finally, the best fit obtained while only
an accretion disc was considered. As we see in Figure 10, when the
disc is included, the agreement between the observed data and the
theoretical light curve is quite good. It should be mentioned that
the solution derived in this way is not a unique one. of course, it
shows that the fit is remarkably improved by assuming a disc of some
reliable dimensions and physical parameters.

The parameters of the disc are given in Table 1. As we see, the
shape of the disc was found to be a slab cylinder with half of the
thickness (\emph{adisc}) 1.2 R$_\odot$. The inner and outer radius
of the disc (\emph{rindc} and \emph{routdc}, respectively) were
obtained 3.5 and 8.0 R$_\odot$. Therefore, the inner edge of the
disc is 0.99 R$_\odot$ away from the surface of the primary star,
since the radius of the primary star was obtained 2.51 R$_{\odot}$
(Ghoreyshi et al. 2011). Moreover, on the basis of the values 4.25
and 13.32 R$_{\odot}$ for the radius of the secondary star and the
distance between the center of the stars, respectively (Ghoreyshi et
al. 2011), it is clear that the outer edge of the disc is 1.07
R$_{\odot}$ away from the surface of the secondary star.

Effective temperature of the disc (\emph{tempdc}) was obtained 5700
K. It is close to that of for the primary star (6000 K). Therefore,
in this system, disc treats as a hot object (in comparison to the
primary star) and totally, increases the overall luminosity of the
system . This is the reason that in Figure 10 we see the system with
disc has more luminosity than that of without disc (against what was
described about Figure 1, since in latter case the effective
temperature of the primary star and disc were assumed to be 9800 and
7000 K, respectively. So, the disc cannot be a hot object. Also, the
overall volume of the disc supposed in Figure 1 is bigger than that
of for this real case and with such temperature, the disc treats as
a cold object in the system and decreases its overall luminosity.
All these points help to justify the difference between the Figure 1
and Figure 10.)

Mean density of the disc (\emph{densdc}) and the electron number
density of the disc (\emph{anedc}) were obtained $33\times10^{-15}$
gr/cm$^3$ and $21\times10^9$ cm$^{-3}$, respectively. Finally, the
best values for the exponent of the power-law behavior of the
densities $\rho\sim$ $r^\eta$ in the disc (\emph{edendc}) and
microturbulence ($v_{trb}$) were obtained for -1 and 90 km/s,
respectively.

The configuration of the system while its primary star is surrounded
by the accretion disc on the basis of the values of Table 1 are
shown in Figure 11.

\begin{table}[t]
\begin{center}
\caption{The parameters of the accretion disc of AV Del}
\begin{tabular}{@{}ll}
\hline & \\
Parameter & Value \\
& \\
\hline & \\
$adisc$ (R$_\odot)$ & 1.2 \\

$rindc$ (R$_\odot)$ & 3.5 \\

$routdc$ (R$_\odot)$ & 8.0 \\

$tempdc$ (K)& 5700 \\

$densdc$ (gr/cm$^3)$ & $33\times10^{-15}$ \\

$anedc$ (cm$^{-3})$ & $21\times10^9$ \\

$edendc$ & -1 \\

$vtrbdc$ (km/s) & 90 \\

\hline
\end{tabular}
\end{center}
\end{table}


\section{Conclusion}

We have used a new computer code called SHELLSPEC for studying the
effect of various free parameters of circumstellar material on the
light curve of a typical Algol-type eclipsing binary system. It is
for the first time that the effect of such parameters are
considered. The temperature and inclination of the disc have the
strongest effect on the emerging light curve of the system, while
the inner and outer radius of the disc have the poorest effect on
the light curve. This is in agreement to what Budaj et al. (2005)
had derived about the spectrum of the TT Hya.

The code was applied to the well-known eclipsing Algol-type binary
system AV Del, which has an accretion disc. Light curve of the
system for all phases were calculated by taking into account a
spherical primary and a Roche lobe–filling secondary, as well as a
disc. It was showed that without an accretion disc the light curve
of the system cannot be described. Therefore, we believe this study
can be a verification for the SHELLSPEC code to analysis of the
eclipsing binary systems, specially those who are doubtful about
existing circumstellar material in their structure. Also, it can be
as a suggestion for future investigation of such systems that the
SHELLSPEC code be used for analysis of their light curve to consider
whether they contain any kind of circumstellar material or not. Of
course, as an important weakness for the SHELLSPEC code in
comparison with other codes like Wilson-Devinney code, we can
mention to inability of the SHELLSPEC code to drive the uncertainty
for the parameters so that one does not know how reliable the values
obtained for the parameters are. Otherwise, the SHELLSPEC code
really is a strong code to analysis of the eclipsing binary system.
\begin{figure*}
\begin{center}
\includegraphics[width=0.6\textwidth]{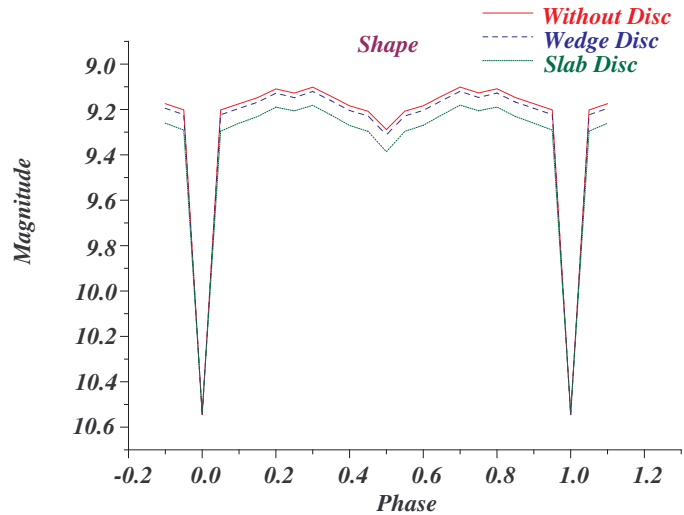}
\caption{Effect of the geometrical shape of the disc on the emerging
light curve of a typical binary system.}
\end{center}
\end{figure*}

\begin{figure*}
\includegraphics[width=1.0\textwidth]{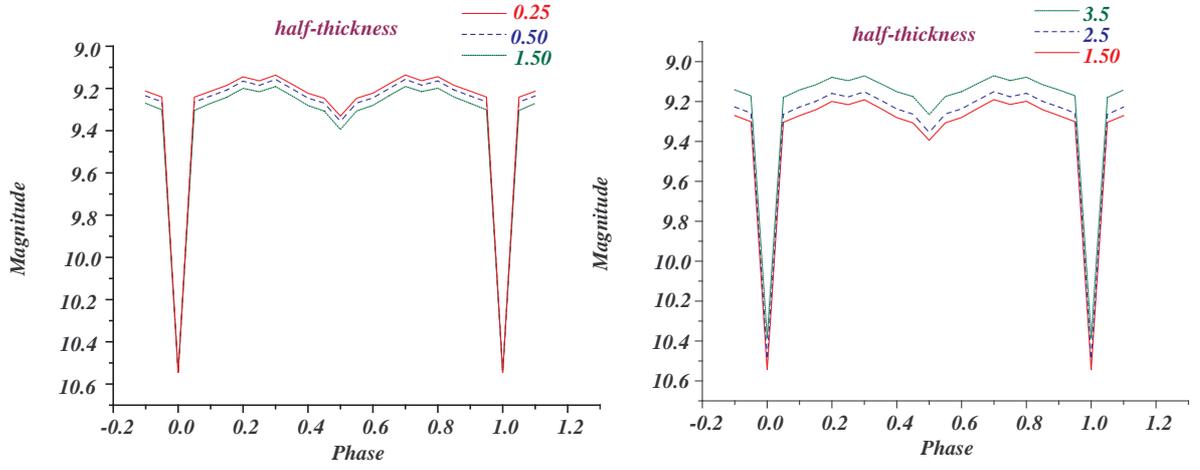}
\caption{$\emph{left}$: Effect of varying $\alpha$ (vertical
half-width of the disc) from 0.25 to 1.5  R$_\odot$; $\emph{right}$:
effect of varying vertical half-width of the disc from 1.5 to 3.5
R$_\odot$.}
\end{figure*}

\begin{figure*}
\begin{center}
\includegraphics[width=0.6\textwidth]{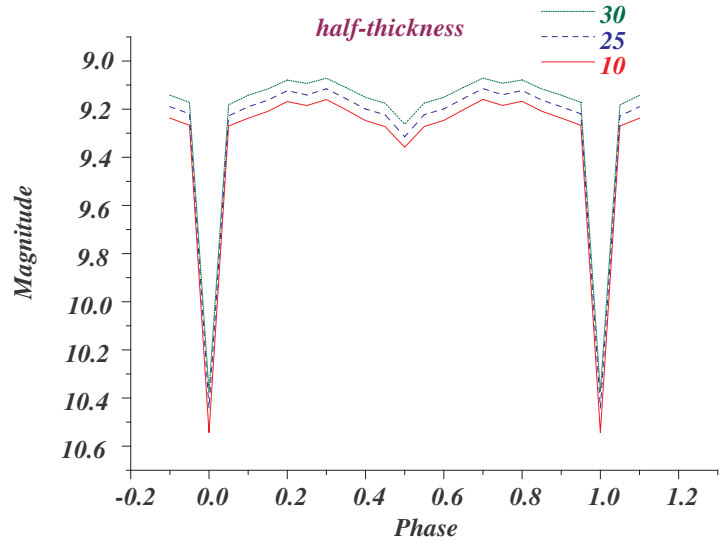}
\caption{Effect of varying the vertical half-width of the disc from
10$^\circ$ to 30$^\circ$ for the rotating wedge disc model.}
\end{center}
\end{figure*}

\begin{figure*}
\includegraphics[width=1.0\textwidth]{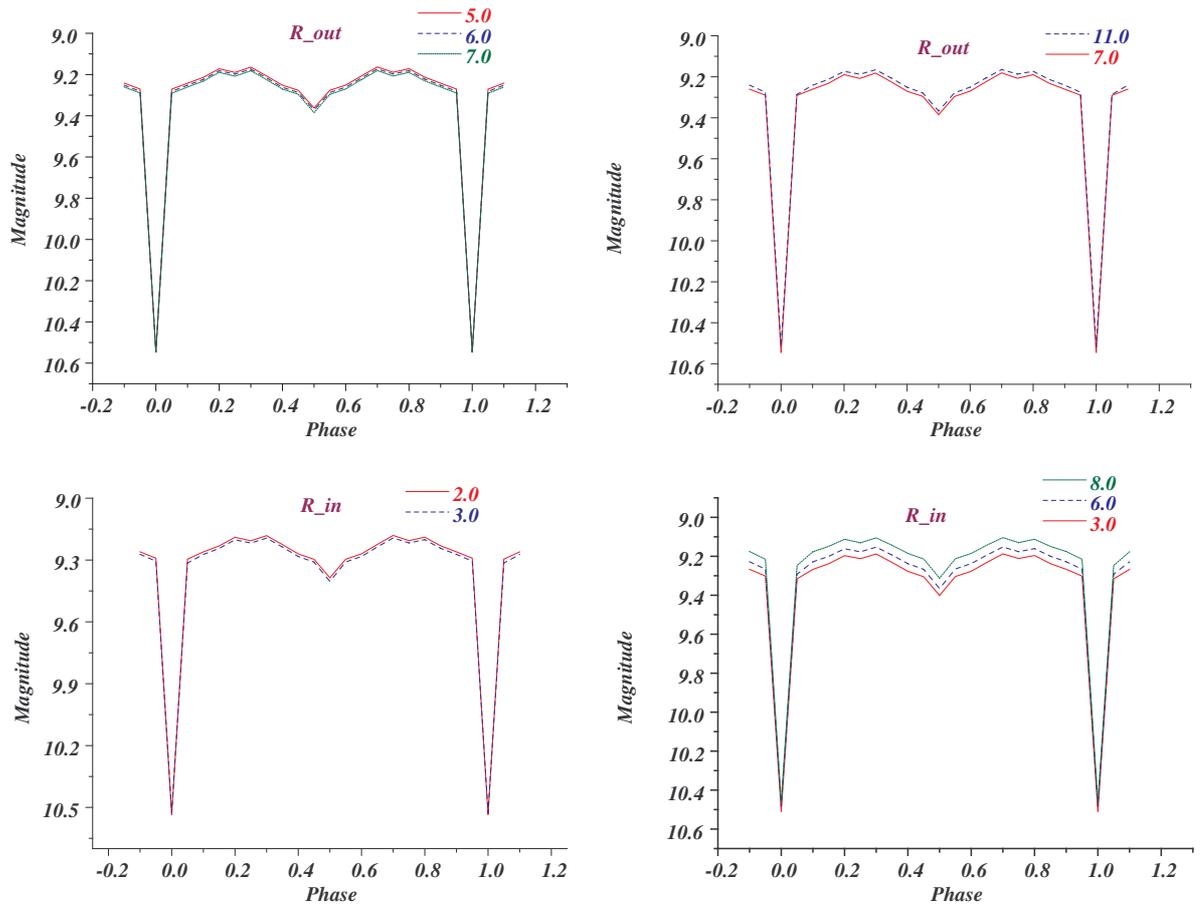}
\caption{$\emph{Top left}$: Effect of varying the outer radius of
the disc from 5 to 7 R$_\odot$; $\emph{top right}$: effect of
varying the outer radius of the disc from 7 to 11 R$_\odot$;
$\emph{bottom left}$: effect of varying the inner radius of the disc
from 2 to 3 R$_\odot$; $\emph{bottom right}$: effect of varying the
inner radius of the disc from 3 to 8 R$_\odot$.}
\end{figure*}

\begin{figure*}
\begin{center}
\includegraphics[width=0.6\textwidth]{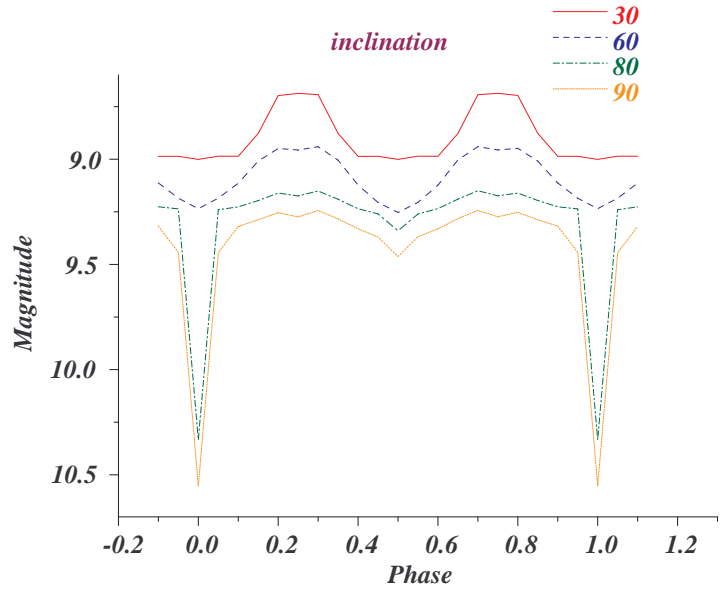}
\caption{Effect of changing the inclination of the disc from
90$^\circ$ to 30$^\circ$.}
\end{center}
\end{figure*}


\begin{figure*}
\begin{center}
\includegraphics[width=1.0\textwidth]{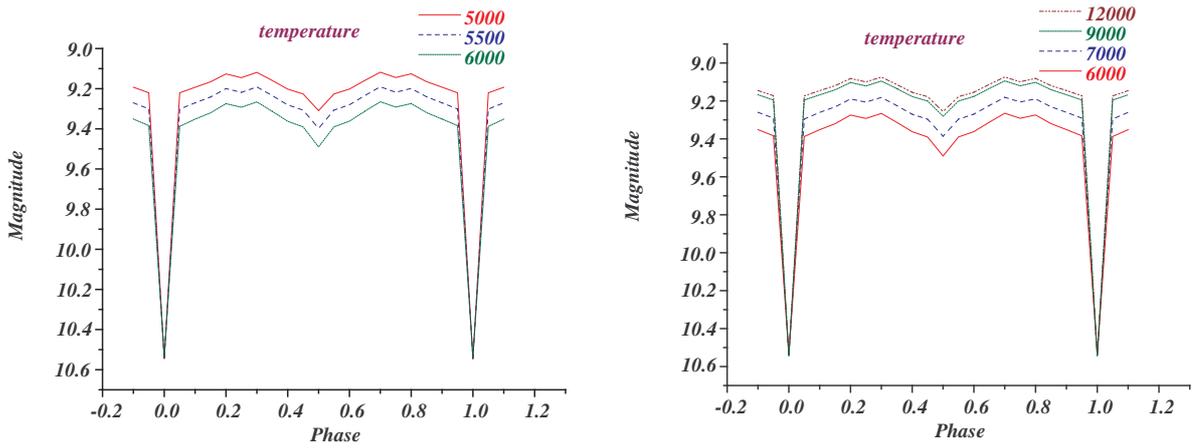}
\caption{$\emph{left}$: Effect of varying temperature of the disc
from 5000 to 6000 K; $\emph{right}$: effect of varying temperature
of the disc from 6000 to 12000 K.}
\end{center}
\end{figure*}

\begin{figure*}
\begin{center}
\includegraphics[width=0.60\textwidth]{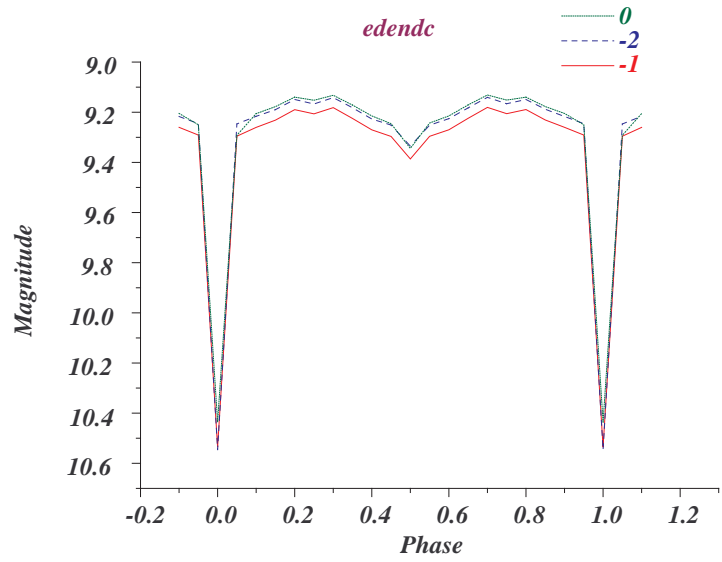}
\caption{Effect of varying the exponent $\eta$ of the power-law
behavior of the densities $\rho\sim r^\eta$ in the disc from 0 to
-2.}
\end{center}
\end{figure*}


\begin{figure*}
\begin{center}
\includegraphics[width=0.60\textwidth]{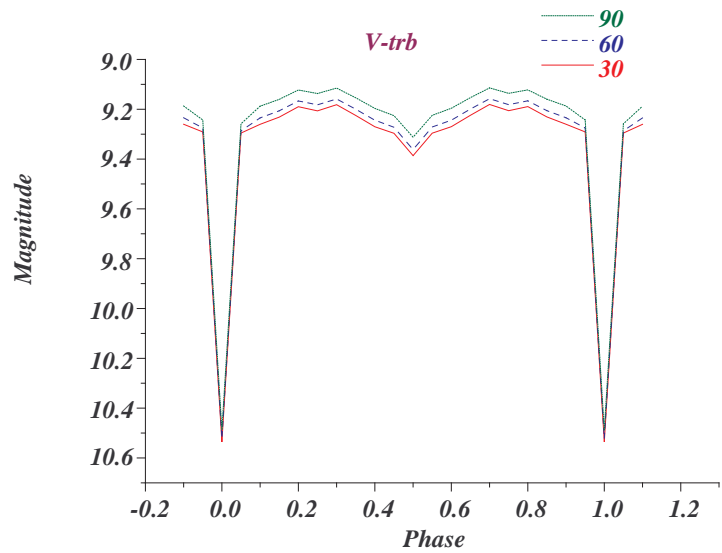}
\caption{Effect of increasing the microturbulence in the disc from
30 to 90 kms$^{-1}$.}
\end{center}
\end{figure*}


\begin{figure*}
\begin{center}
\includegraphics[width=1.0\textwidth]{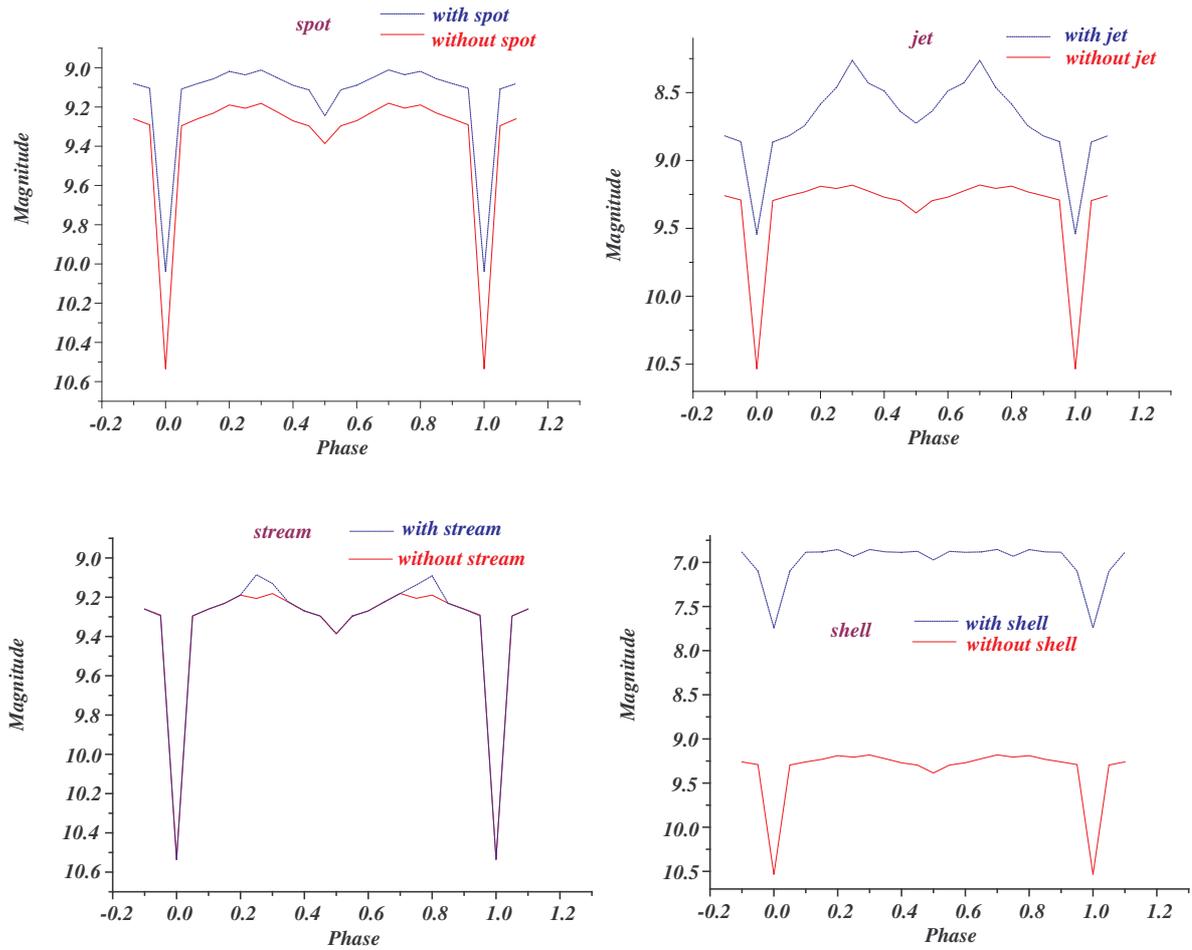}
\caption{$\emph{Top left}$: Effect of existence of the spot on the
emerging light curve of a typical binary system; $\emph{top right}$:
effect of existence of the jet on the emerging light curve of a
typical binary system; $\emph{bottom left}$: effect of existence of
the stream on the emerging light curve of a typical binary system;
$\emph{bottom right}$: effect of existence of the shell on the
emerging light curve of a typical binary system.}
\end{center}
\end{figure*}


\begin{figure*}
\begin{center}
\includegraphics[width=0.60\textwidth]{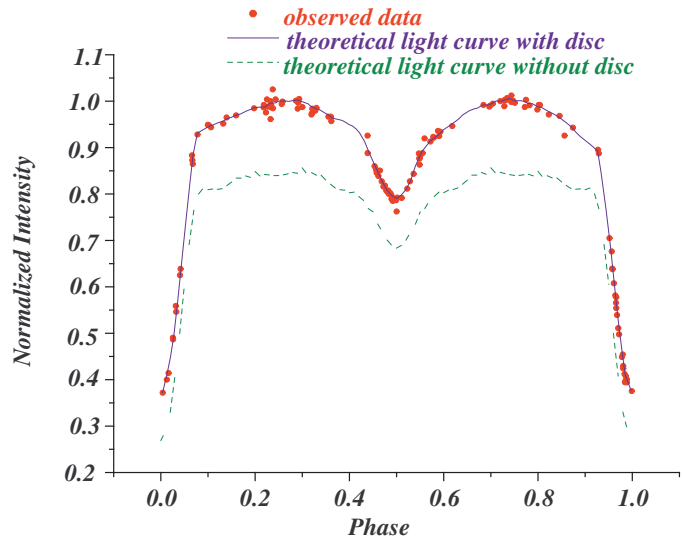}
\caption{The observed and the theoretical light curves of AV Del.
The theoretical light curves are drawn on the basis of the values
derived via SHELLSPEC code. The observed data are shown by solid
circles and the theoretical light curves are shown by continuous
(with disc) and dashed (without disc) lines.}
\end{center}
\end{figure*}


\begin{figure*}
\begin{center}
\includegraphics[width=0.80\textwidth]{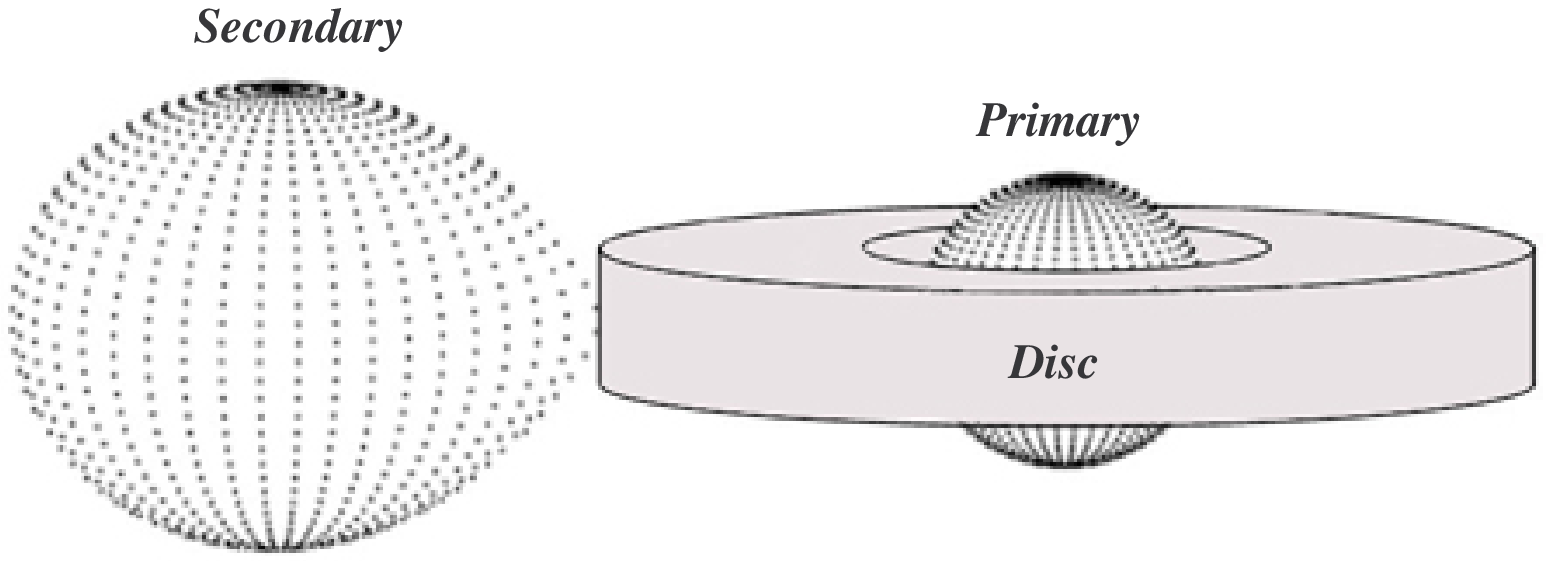}
\caption{The Configuration of the binary system AV Del on the basis
of the values derived by using the SHELLSPEC code. As we see, the
primary star is surrounded by an accretion disc.}
\end{center}
\end{figure*}


\section*{Acknowledgments} We wish to thank Dr. Jano Budaj for his very useful comments at
different stages of the work. Also we would like to thank the
referee for his (her) comments and suggestions.


\end{document}